\documentclass[%
 reprint,
 superscriptaddress,
 amsmath,amssymb,
 aps,
]{revtex4-2}

\usepackage{xcolor}
\usepackage{graphicx}
\usepackage{dcolumn}
\usepackage{bm}
\usepackage[colorlinks=true,linkcolor=blue,citecolor=blue,urlcolor=blue]{hyperref} 

\DeclareMathOperator*{\argmin}{argmin}

\begin{document}

\preprint{APS/123-QED}

\title{Linear scaling causal discovery from high-dimensional time series by dynamical community detection}

\author{Matteo Allione}\thanks{These two authors contributed equally to this work.}
\affiliation{Scuola Internazionale Superiore di Studi Avanzati (SISSA), Via Bonomea 265, 34136 Trieste, Italy}

\author{Vittorio Del Tatto}\thanks{These two authors contributed equally to this work.}
\affiliation{Scuola Internazionale Superiore di Studi Avanzati (SISSA), Via Bonomea 265, 34136 Trieste, Italy}

\author{Alessandro Laio}
\email{Contact author: laio@sissa.it}
\affiliation{Scuola Internazionale Superiore di Studi Avanzati (SISSA), Via Bonomea 265, 34136 Trieste, Italy}
\affiliation{International Centre for Theoretical Physics (ICTP), Strada Costiera 11, 34151 Trieste, Italy}


\begin{abstract}

Understanding which parts of a dynamical system cause each other is extremely relevant in fundamental and applied sciences.
However, inferring causal links from observational data, namely without direct manipulations of the system, is still  computationally challenging, especially if the data are high-dimensional.
In this study we introduce a framework for constructing causal graphs from high-dimensional time series, whose computational cost scales linearly with the number of variables. 
The approach is based on the automatic identification of \emph{dynamical communities}, groups of variables which mutually influence each other and can therefore be described as a single node in a causal graph. 
These communities are efficiently identified by optimizing the Information Imbalance, a statistical quantity that assigns a weight to each putative causal variable based on its information content relative to a target variable.
The communities are then ordered starting from the fully autonomous ones, whose evolution is independent from all the others, to those that are progressively dependent on  other communities, building in this manner a community causal graph.
We demonstrate the computational efficiency and the accuracy of our approach on discrete-time and continuous-time dynamical systems including up to 80 variables.

\end{abstract}

\maketitle



\textit{Introduction} --- 
The growing abundance of time series - spanning fields from environmental monitoring to finance and neuroscience - offers unprecedented opportunities for understanding real world observations. 
Specifically, causal discovery allows studying how different parts of a system influence each other and inferring the existence of directional couplings (causal relationships) between variables in a time series \cite{runge2023causal_inference,runge2018causal_network_reconstruction,assaad2022survey}. 

Relationships between variables can be depicted, using Pearl's approach \cite{pearl2009causality, spirtes2001causation}, as a time series graph. 
Each dynamic variable at a specific time is represented by a node and an arrow from one node to another represents a direct causal link. 
Such a graph 
encodes all conditional independence relationships between pairs of variables \cite{geiger1990identifying,verma1990causal,runge2018causal_network_reconstruction}.
Several strategies can be used to build this graph. 
A first option involves checking for each couple of lagged variables if there is no conditioning set that makes them independent. 
An arrow between the two nodes is then drawn only if such set does not exist~\cite{chicharro2014algorithms_causal_inference}.
A strategy to assess this condition is to employ an iterative approach, where the size of the tested set is progressively increased \cite{verma1990equivalence_IC,spirtes1991analgorithm_PC}.
Although this approach leads to optimal detection power,
it requires performing a number of tests that scales exponentially with the number of variables. 
Therefore, strategies to reduce the search space in practical applications have been developed \cite{runge2019detecting_quantifying,runge2020discovering_pcmciplus}.
Alternatively,
one can consider for each pair of lagged variables the largest possible conditioning set, which includes all past history of the time series.
The multivariate versions of Granger Causality~\cite{granger1969investigating,barrett2010multivariate_GC,blinowska2004granger} and Transfer Entropy \cite{schreiber2000measuring} can be regarded as implementations of this second strategy \cite{chicharro2014algorithms_causal_inference, runge2018causal_network_reconstruction}.
Although this approach drastically reduces the number of tests to be performed, requiring a single test per pair of variables, it is significantly affected by the curse of dimensionality, often resulting in low detection power \cite{runge2019detecting_quantifying}.

In this work, we introduce a method for causal discovery designed for high-dimensional time series. 
A powerful feature of our approach is that it enables the identification of causal influences that emerge from the collective dynamics of multiple variables at a very moderate computational cost, which, importantly, scales linearly with the number of variables.
The graph obtained with our algorithm is a ``mesoscopic'' version of the standard graphs, as it groups together variables whose evolution cannot be described independently. We call these groups of variables \emph{dynamical communities}. Our approach aims at revealing, if present, a hierarchical organization of these communities, emerging from unidirectional inter-group interactions.  We refer to the resulting graph as \emph{community causal graph}. 

The key ingredient of our method is the efficient identification of the dynamical communities. 
This is achieved by optimizing the Information Imbalance \cite{glielmo2022ranking,wild2025automatic} as a function of a set of variational weights, one for each variable of the system, according to a prediction criterion broadly inspired by Granger Causality~\cite{barrett2010multivariate_GC,blinowska2004granger}. 
We will show that the values of the weights allow identifying efficiently the communities and, as a consequence, building a causal graph, avoiding combinatorial searches of conditioning sets.

We build the method  on the assumption of causal sufficiency, namely that there are no unobserved common drivers of two or more dynamic variables. In the Supplemental Material (SM) we discuss the effect of unobserved variables in different scenarios.
We demonstrate the effectiveness of our approach using time series generated from both discrete-time and continuous dynamical systems, showcasing its applicability to high-dimensional scenarios.


The algorithm exploits the Information Imbalance \cite{glielmo2022ranking}, which allows quantifying the information content of different distance measures defined on a data set. 
The underlying idea is that a distance measure $d^A$ is predictive with respect to another distance measure $d^B$ if points close according to $d_A$ are also close according to $d^B$.
The Information Imbalance is defined as
\begin{equation}\label{eq:std_info_imb}
    \Delta(d^A\rightarrow d^B)=\frac{2}{N^2} \sum_{\underset{(i\neq j)}{i,j}} \delta_{r^A_{ij},1}r^B_{ij}\,,
\end{equation}
where $N$ is the number of points in the dataset, $\delta$ is the Kronecker delta and $r_{ij}^{(\cdot)}$ is the distance rank of point $j$ with respect to point $i$.
The superscript refers to the distance used in the computation.
For example, $r_{ij}^A = 2$ if $j$ is the second nearest neighbor of $i$ according to distance $d^A$.
Eq.~(\ref{eq:std_info_imb}) defines a quantity that, in the limit of large $N$, approaches zero when all nearest neighbors in space $A$ remain nearest neighbors in space $B$, namely when $d^A$ is maximally predictive of $d^B$.

In ref.~\cite{deltatto2024robust_inference} we showed that the Information Imbalance can be used to infer the presence of causality between two multidimensional dynamical systems $X$ and $Y$.
We assumed that if $X$ causes $Y$ and one attempts to make a prediction of the future of $Y$, a distance measure built using the present states of both $X$ and $Y$ will have more predictive power than a distance built using only $Y$.
Formally, we assumed that $X$ causes $Y$ if and only if
\begin{equation}\label{eq:ineq}
    \hat{w} = \argmin_w \Delta\big(d^{w X(0),Y(0)}\rightarrow d^{Y\left(\tau\right)} \big) \neq 0
\end{equation}
for some positive time lag $\tau$.
The notation $d^{(\cdot)}$ denotes the squared Euclidean distance built over the superscript variables.
For example, 
\begin{equation}\label{eq:euclidean_dist_ij}
    d_{\,ij}^{w X(0),Y(0)} = w^2 \| X_i(0) - X_j(0) \|^2 + \| Y_i(0) - Y_j(0)\|^2\,,
\end{equation}
where the Latin letters $i$ and $j$ denote independent realizations of the same dynamics, obtained either from uncorrelated samples of a single stationary trajectory, or from distinct trajectories with independent initial conditions.
In the SM we relate the criterion in Eq.~(\ref{eq:ineq}) to the notion of conditional independence. 
Importantly, the approach in ref.~\cite{deltatto2024robust_inference} builds on prior knowledge of the groups of variables that make up the distinct dynamical systems interacting with each other (the dynamical communities in the language of this work, see below).
This is a very strong assumption, which for real-world data is typically violated. This work is dedicated to overcoming this problem.

Here, we extend the approach described above to automatically and efficiently find those dynamical communities.
Our algorithm makes use of a differentiable extension of the Information Imbalance (DII) \cite{wild2025automatic}:
\begin{equation}\label{eq:diff_imb}
    \text{DII}\,(d^A\rightarrow d^B) = \frac{2}{N^2}\sum_{\underset{(i\neq j)}{i,j}} \frac{e^{-d^A_{ij}/\lambda}}{\sum_{m(\neq i)}e^{-d^A_{im}/\lambda}}r_{ij}^B\,.
\end{equation}
We note that Eq.(\ref{eq:diff_imb}) tends to Eq.~(\ref{eq:std_info_imb}) in the limit $\lambda~\rightarrow~0$.
If distance $d^A$ depends on a set of parameters $\boldsymbol{w}$, this formulation allows optimizing such parameters by gradient descent. 

\begin{figure*}
    \centering
    \includegraphics[width=.95\textwidth]{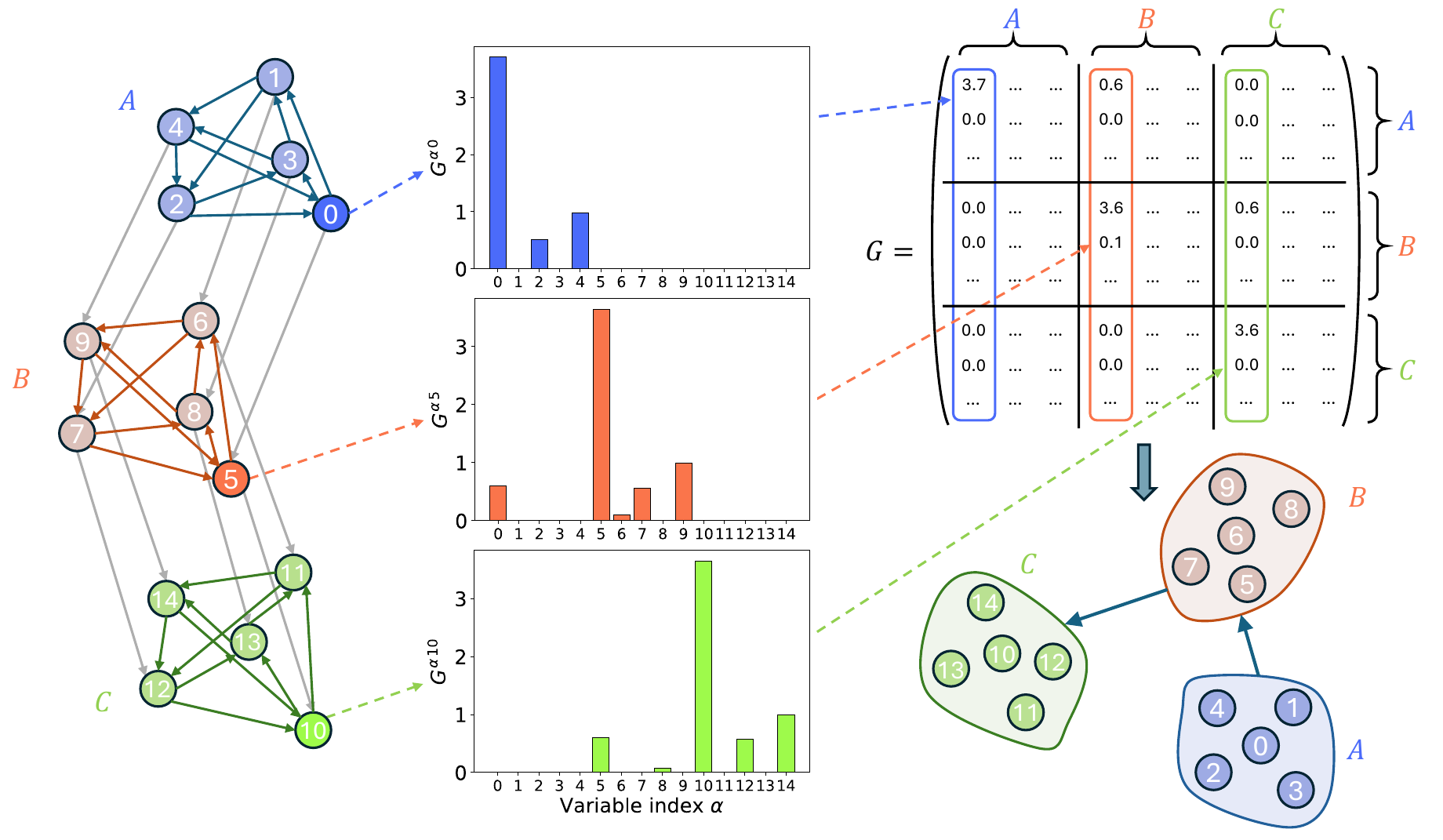}
    \caption{
    Illustration of the algorithm, using a 15-dimensional dynamical system composed of three groups of noisy coupled logistic maps.
    On the left, the ground-truth connectivity of the system is depicted in an all-variable representation.
    Repeated optimizations of DII$( d^{\boldsymbol{w}\odot \boldsymbol{X}(0)} \rightarrow d^{X^\beta(\tau)})$ are carried out for each target variable $\beta$ according to Eq.~(\ref{eq:argmin_w}) ($\beta = 0, 5$ and 10 are shown in blue, red and green, respectively).
    The optimal weights from each optimization, depicted in the central bar plots, are used to construct a connectivity matrix $G$ (here, $G$ is built using a single time lag $\tau = 1$).
    From this matrix, dynamical communities are identified and depicted on a graph, where several variables are grouped within the same node.}
    \label{fig:framework}
\end{figure*}

Our algorithm, illustrated in Fig. \ref{fig:framework}, can be conceptually divided into three parts: in part i) the DII is minimized to infer each variable’s \emph{autonomous set}, namely the set of variables that directly or indirectly cause it;
in part ii), the autonomous sets are used to find the dynamical communities, namely sets of variables which directly or indirectly influence each other; and in part iii), a macroscopic graph depicting the causal interactions between the communities is constructed.
The input to the algorithm is a set of $D$ time-dependent variables $\{X^\alpha(t)\}_{\alpha=1}^D$. 
We will refer to such variables as ``microscopic'', in contrast to the ``mesoscopic'' dynamical communities identified by the method.

\textit{i) Identification of autonomous sets} --- As a first step we infer the \textit{autonomous set} $\mathcal{S}^\beta$ of each variable $X^\beta$, which we define as the set of all variables $\{X^\alpha\}$ directly or indirectly causing $X^\beta$.
$X^\alpha$ is a direct cause of $X^\beta$ if, for some time lag $\tau$, there exists a direct link $X^\alpha(0)\rightarrow X^\beta(\tau)$ in the ground-truth time series graph.
Conversely, $X^\alpha$ is an indirect cause of $X^\beta$ if, for any $\tau > 0$, the directed paths connecting $X^\alpha(0)$ and $ X^\beta(\tau)$ pass through at least a third variable $X^\gamma(\tau')$ ($\gamma \neq \alpha$, $\gamma \neq \beta$, $0 < \tau' < \tau$).
We will use the notations $X^\alpha\rightarrow X^\beta$ and $X^\alpha\in\mathcal{S}^\beta$ interchangeably, without distinguishing between direct and indirect links among the microscopic variables.

To identify the autonomous sets $\mathcal{S}^\beta$, we optimize the Information Imbalance of Eq.~(\ref{eq:diff_imb}) between a distance measure built with all dynamic variables at time $t=0$ and a distance built with a single variable $X^\beta$ at time $t=\tau$.
To level out the fluctuation ranges of different variables, we first scale each variable by its standard deviation over the entire trajectory.
Thus, the weights obtained by this optimization are 
\begin{equation}\label{eq:argmin_w}
    \hat{\boldsymbol{w}}_{\beta}
    =\argmin_{\boldsymbol{w}} \ \text{DII}\left(  d^{\boldsymbol{w}\odot \boldsymbol{X}(0)} \rightarrow d^{X^\beta(\tau)} \right)\,, 
\end{equation}
where $\odot$ denotes the element-wise product.
We generalize the principle of Eq.~(\ref{eq:ineq}) by stating that $X^\alpha$ is a direct or indirect cause of $X^\beta$ when the $\alpha$-component of $\hat{\boldsymbol{w}}_\beta$, denoted by $\hat{w}_\beta^\alpha$, is nonzero.

Following the intuition that different couplings might manifest at different time scales \cite{Tirabassi2015-wy, Smirnov_time_scales, deltatto2024robust_inference} we repeat the optimization in Eq.~(\ref{eq:argmin_w}) for several values of $\tau$ between 1 and $\tau_{\text{max}}$, where $\tau_{\text{max}}$ is a hyper parameter which, in applications, we take of the order of the autocorrelation time of $X^\beta$.
As depicted in Fig.~\ref{fig:framework}, the maximum weights over the tested values of $\tau$ are stored as columns of a $D \times D $ matrix $G$:
\begin{equation}\label{eq:matrix_G}
    G^{\alpha\beta} = \max_{\tau} \hat{w}_\beta^\alpha\,.
\end{equation}
Constructing $d^A$ over a single time frame provides the correct results when the lag of direct links is not larger than 1 in the underlying time series graph. In the SM we describe how our approach can be extended when this condition does not hold, constructing $d^A$ on multiple time frames.

Ideally, each autonomous set $\mathcal{S}^\beta$ can be directly extracted from the nonzero elements of $G^{\alpha\beta}$. 
However, in applications $G$ is estimated using a finite number of measures (or a finite trajectory).
Therefore, one can decide if an element is zero only according to a specified tolerance: we then set $X^\alpha\in \mathcal{S}^\beta$ whenever $G^{\alpha\beta} > \varepsilon$.
The threshold $\varepsilon$ is the main hyper parameter of our algorithm.
A false negative may appear if the 
coupling that we aim to detect by testing $\hat{w}_\beta^\alpha >0$
is present but weak.
This may occur, for example, when $X^\alpha$ causes $X^\beta$ indirectly via several mediating variables.

To account for false negatives, we first construct a directed graph represented by the set of links for which $G^{\alpha\beta} > \varepsilon$, and then we construct each autonomous set $\mathcal{S}^\beta$ as the full set of ancestors of $X^\beta$.
Missing links $X^\alpha\rightarrow X^\beta$ are likely to be recovered through indirect paths $X^\alpha\rightarrow X^\gamma \rightarrow ... \rightarrow X^\beta$.

\textit{ii) Identification of dynamical communities} ---  
A hierarchical structure of dependencies between groups of variables can be directly retrieved from the analysis of the sets $\mathcal{S}^\beta$. 

In particular, given an autonomous set $S^\beta$, we will define it as \textit{minimal} if, for every variable $x^\alpha \in S^\beta$, one has $S^\alpha \equiv S^\beta$. 
In a minimal autonomous set, each variable depends on all and only the other variables in the set: together they form a \textit{dynamical community} $\mathcal{G}_k$ whose evolution is independent of the rest of the network.
Once all sets $\mathcal{G}_k$ have been identified, 
their variables can be removed from the graph, and 
new sets with the same property can be extracted.
This will reveal new dynamical communities, which we distinguish from the initial ones by assigning them a higher autonomy level.
By convention, we assign autonomy level 0 to the first dynamical communities identified, and we increase such a level by 1 at each iteration. 
Notice that, if a group has level of autonomy $k$, then it can be caused
only by groups with level $m < k$, and, in turn, it can cause only groups with level $m > k$; no links can be present among groups at the same level. 

\begin{figure*}[ht!]
    \centering
    \includegraphics[width=.95\textwidth]{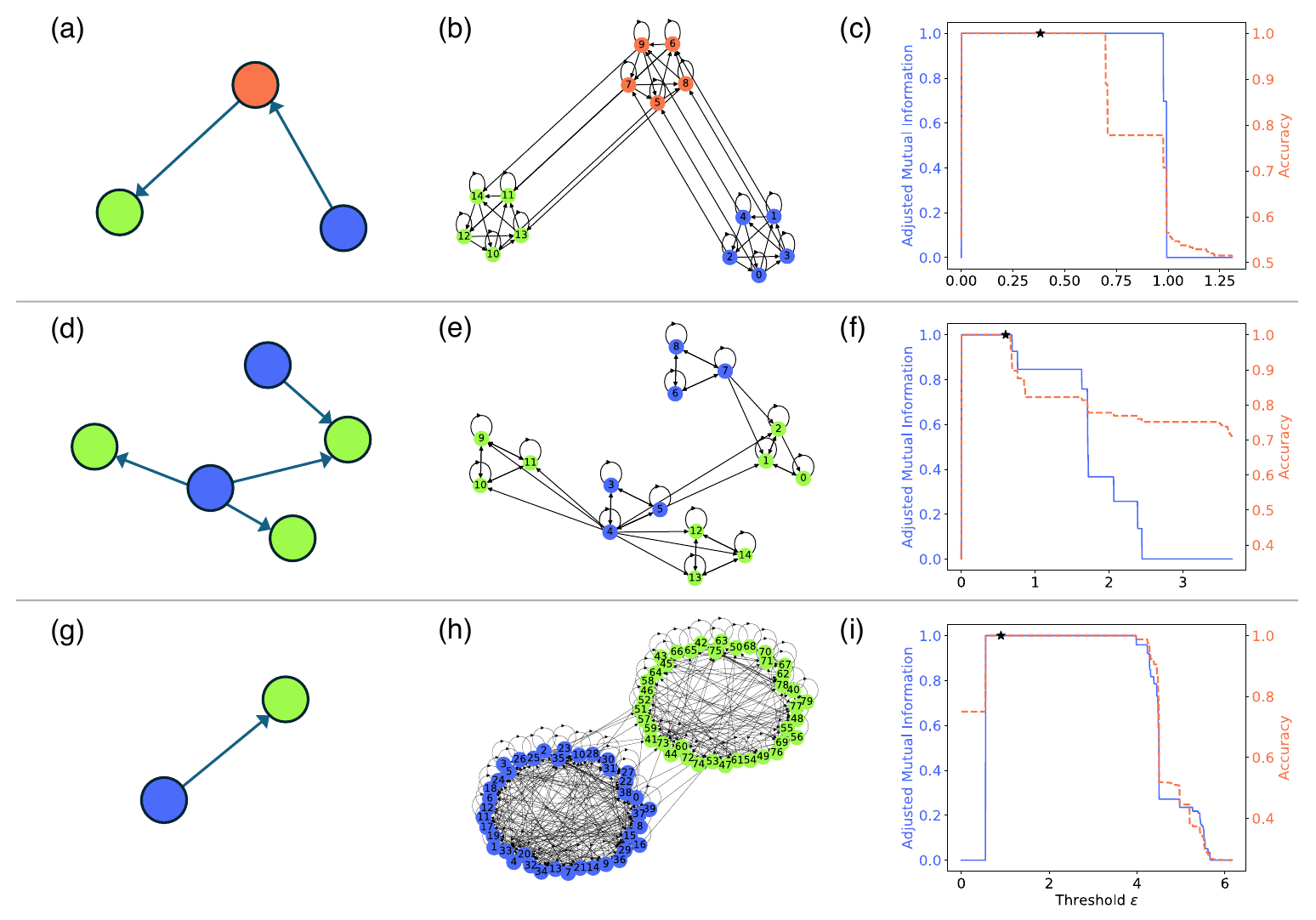}
    \caption{Outcome and performance of the algorithm in three different test cases: (a)-(c) 15 coupled logistic maps, (d)-(f) 5 coupled Lorenz systems and (g)-(i) 2 Lorenz 96 systems.
    Left panels (a),(d) and (g): community causal graphs produced by the algorithm, which correctly retrieves the ground-truth connectivity.
    Central panels (b), (e) and (h): all-variable graphs obtained from condition $G^{\alpha\beta} > \varepsilon$, setting $\varepsilon$ to 0.38, 0.6 and 3.5, respectively.
    Nodes in community and all-variable graphs are colored according to the autonomous levels identified in step ii).
    Right panels (c), (f) and (i): validation measures of the recovered connectivity as a function of the threshold parameter $\varepsilon$, in the range $\left[0, \max_{\alpha \beta\, (\alpha\neq \beta)} G^{\alpha\beta}\right]$.
    For larger values of $\varepsilon$, no pairs of variables are found to be linked.
    Blue solid curve: adjusted mutual information between the retrieved groups and the ground-truth ones.
    Red dashed curve: accuracy of the all-variable adjacency matrix recovered from the final community graph, defined as the fraction of correctly identified links.
    Black star marker: results by setting the threshold $\varepsilon$ to the average of all weights in $G^{\alpha\beta}$.
    }
    \label{fig:results}
\end{figure*}



\textit{iii) Construction of the community causal graph} --- The causal connections between the dynamical communities are depicted using a directed acyclic graph where each node represents a dynamical community $\mathcal{G}_k$, which we name \textit{community causal graph} (see Fig.~\ref{fig:framework}).
We draw a directed arrow $\mathcal{G}_k \rightarrow \mathcal{G}_m$ if there is at least a pair of variables $X^\alpha\in\mathcal{G}_k$ and $X^\beta\in\mathcal{G}_m$ such that $X^\alpha \rightarrow X^\beta$. Directed arrows $\mathcal{G}_k \rightarrow \mathcal{G}_m$ are omitted whenever $\mathcal{G}_k$ and $\mathcal{G}_m$ are also connected through an indirect path $\mathcal{G}_k \rightarrow ... \rightarrow \mathcal{G}_m$.

By construction, links that connect communities with consecutive levels of autonomy are direct (e.g., $\mathcal{G}^k\rightarrow \mathcal{G}^m$), whereas links between non-consecutive communities may be indirect.
In \emph{Appendix A}, we show that one can  distinguish between direct and indirect links among non-consecutive communities by a simple \emph{a posteriori} analysis.

\textit{Results} --- We tested our approach on trajectories generated by dynamical systems of different complexity.
Specifically, we considered three groups of five coupled logistic maps (Figs.~\ref{fig:results}a-c), five coupled Lorenz oscillators \cite{lorenz1963deterministic} (Figs.~\ref{fig:results}d-f), and  two 40-dimensional Lorenz 96 systems \cite{lorenz2006predictability} unidirectionally coupled (Figs.~\ref{fig:results}g-i).
Since deterministic relationships are known to violate a condition known as faithfulness \cite{runge2018causal_network_reconstruction,chicharro2014algorithms_causal_inference}
we added a small white noise to each variable while integrating the dynamic equations (see \emph{Appendix B}).
We report in the SM additional tests in presence of observational noise.
For each system, distances entering the DII optimization at step i) of the algorithm were computed by extracting $N=2000$ frames, used as independent initial conditions, from a single time series realization.

The left panels in Fig.~\ref{fig:results} show the community graphs produced by the algorithm when the threshold $\varepsilon$ is set to the average value of the weights in matrix $G$ (black star marker in the right panels).
These graphs reproduce the correct connectivity read from the ground-truth equations: all variables grouped in a single node are dynamically intertwined, namely each one is a direct or indirect cause of any other variable, and an arrow between two groups at consecutive levels is present when at least two variables, one per group, are interacting.
The central panels show the all-variable graphs built from the matrix $G^{\alpha\beta} > \varepsilon$, with nodes colored according to the level of autonomy assigned to the corresponding groups at step ii).
This matrix was computed according to Eq.~(\ref{eq:matrix_G}), with $\tau_{\text{max}}=1$ for the top row example, $\tau_{\text{max}}=60$ for the middle row, and $\tau_{\text{max}}=30$ for the bottom row example.

The outcome of the algorithm is influenced by the choice of the threshold $\varepsilon$.
In Fig.~\ref{fig:results}, right column, we validated the correctness of the recovered dynamical communities and their connectivity separately, by measuring as a function of $\varepsilon$ the adjusted mutual information \cite{vinh2009information_AMI} between the true groups and the recollected ones (right panels, blue solid curves), and the accuracy of the all-variable adjacency matrix obtained from the final community graphs (right panels, dashed red curves).
Both measures are defined in the range $[0,1]$, with 1 being the case of optimal reconstruction.
We refer to the SM for a formal definition of these measures.
Remarkably, in all systems we observe a wide range of threshold values leading to an exact reconstruction, and a false positive rate which is nonzero only for negligibly small values of the threshold ($\epsilon < 10^{-16}$, $\epsilon < 10^{-13}$ and $\epsilon < 0.55$ for the systems in panels a, d and g, respectively).

\textit{Discussion} --- We have demonstrated that our approach can efficiently reconstruct the causal structure underlying high-dimensional dynamical systems, providing a coarse-grained visualization of the system's causal connectivity, which has been a topic of growing interest \cite{wahl2024foundations}.
The most relevant feature of our method is that the number of optimizations — analogous to conditional independence tests — scales linearly with the number of variables, and still the approach is able to capture ``multibody'' synergistic causal effects that are hard to detect by constructing conditioning sets of increasing size \cite{runge2018causal_network_reconstruction}.

For comparison, we applied a state-of-the-art method for causal discovery on time series data, PCMCI \cite{runge2019detecting_quantifying}, to the coupled Lorenz 96 systems.
This algorithm efficiently reduces the search space for conditioning sets, at the cost of introducing some hyperparameters.
In the SM, we show that the two 40-dimensional communities are correctly identified for certain hyperparameter combinations.
However, the link among the communities appears hard to detect, as it results from few inter-community connections. 


We notice one could consider analyzing the G matrix 
with graph clustering methods \cite{SCHAEFFER200727}, rather than using it to identify communities. One could also employ this matrix to speed up the search of relevant groups of variables according to other criteria \cite{tononi1998functional,villani2015search_candidate,daddese2021asymptotic,daddese2021fast_effective_method}, even in presence of bidirectional couplings.
The approach can be improved by estimating the statistical confidence of each element in $G$, allowing the application of confidence-based thresholds, instead of a fixed $\varepsilon$. We consider the usage of multiple time delays $\tau$ in Eq.~(\ref{eq:matrix_G}) as a valuable feature of our approach. In the SM (Fig.~S2) we show that using only $\tau = 1$ would significantly decrease the detection power of the algorithm.

The codes implementing our algorithm are available in the Python library DADApy \cite{dadapy2022}.





\textit{Acknowledgments} --- This work was partially funded by NextGenerationEU through the Italian National Centre for HPC (Grant No. CN00000013).
A.L. also acknowledges financial support by the region Friuli Venezia Giulia (project F53C22001770002 received by A.L.).

\bibliography{apssamp}

\newpage
\onecolumngrid
\section*{End Matter}
\twocolumngrid

\emph{Appendix A: Community graph refinement} --- To properly interpret the arrows of the community graph, we define a link between two dynamical communities $\mathcal{A}$ and $\mathcal{B}$ as ``direct'' if there is at least a pair of variables, $X^\alpha \in \mathcal{A}$ and $X^\beta \in \mathcal{B}$, such that $X^\alpha$ is a direct cause of $X^\beta$.
In contrast, we call such a link ``indirect'' if \emph{all} causal paths between their constituent microscopic variables are indirect.
Links that connect communities with consecutive levels of autonomy are necessarily direct, as the absence of ``mediating'' communities implies the existence of at least a pair of microscopic variables, one for each group, that are directly linked.
On the other hand, communities $\mathcal{A}$ and $\mathcal{B}$ in the pattern $\mathcal{A}\rightarrow \mathcal{C} \rightarrow \mathcal{B}$ may be directly linked, although such a link does not correspond to any direct arrow in the community graph.

Whenever a pattern $\mathcal{A} \rightarrow \mathcal{C} \rightarrow \mathcal{B}$ appears in the community graph, the existence of a direct link from $\mathcal{A}$ to $\mathcal{B}$ can be assessed by treating each community as single multi-dimensional variable.
In the following, $\mathcal{C}$ may possibly represent a sequence of mediating groups.
A direct link $\mathcal{A} \rightarrow \mathcal{B}$ is present if there exists a pair of microscopic variables $X^\alpha\in \mathcal{A}$ and $X^\beta\in \mathcal{B}$ that are directly linked.
We note that this definition of inter-group link is consistent with that provided within the general causal discovery framework of ref.~\cite{wahl2024foundations}.
Assuming the absence of instantaneous interactions, this condition is fulfilled if there exists a lag $\tau$ for which $\mathcal{B}(\tau)$ depends on $\mathcal{A}(0)$, conditioning over 
\begin{equation}
    \{ \mathcal{B}(\tau-1),\, ...,\, \mathcal{B}(\tau-E),\, \mathcal{C}(\tau-1),\, …,\, \mathcal{C}(\tau-E) \}\,.
\end{equation}
Here, $E$ represents the maximum lag of the microscopic direct links.
Conditioning over the driven ($\mathcal{B}$) and mediating ($\mathcal{C}$) communities in the $E$ frames preceding $t=\tau$ ensures that all microscopic paths from $\mathcal{A}(0)$ to $\mathcal{B}(\tau)$ are blocked in the sense of $d$-separation~\cite{chicharro2014algorithms_causal_inference}.


Importantly, $\tau$ does not represent the specific lag of the putative direct link.
As commented in the Discussion, the value of $\tau$ that maximizes this conditional dependence in practice may be significantly larger than the maximum lag of the direct microscopic interactions.

The above conditional independence test can be translated into the minimization of the DII, similarly to Eq.~(\ref{eq:argmin_w}).
In the case $E=1$, this minimization reads
\begin{equation}\label{eq:direct_vs_indirect}
    \min_{\boldsymbol{w}}\text{DII}\big(d^{\,\boldsymbol{w} \odot \left[\mathcal{A}(0),\, \mathcal{B}(\tau-1),\,  \mathcal{C}(\tau-1)\right]} 
    \rightarrow d^\mathcal{B(\tau)} \big),
\end{equation}
where $\boldsymbol{w} \odot \left[ \cdot \right]$ denotes a ``group-wise'' product such that, for each community in the list, all its variables are scaled by a single weight (if $\mathcal{C}$ is a single community, $\boldsymbol{w}$ has four components).
If the optimal weight associated to $\mathcal{A}(0)$ is different from zero for some lag $\tau$, we conclude that a direct link $\mathcal{A} \rightarrow \mathcal{B}$ exists, and we draw the corresponding arrow in the refined community graph.

The number of optimizations to be performed 
to test all pairs of linked and non-consecutive communities depends on the topology
of the community graph.
In the worst-scaling scenario, which is the case of a non-branched chain of $D'$ communities ($\mathcal{G}^0\rightarrow \mathcal{G}^1 \rightarrow .... \rightarrow \mathcal{G}^{D'}$), the number of tests is equal to $(D'-1)(D'-2)/2$. This number comes by counting $D'-2-k$ optimizations for each community $\mathcal{G}^k$, where $\mathcal{G}^k$ is the putative ``driver'' community (except for the last two communities, which do not have non-consecutive groups).
We stress that this scaling is quadratic in the number of dynamical communities, which is expected to be substantially smaller than the number of microscopic variables.
Moreover, optimizations such as that of Eq.~(\ref{eq:direct_vs_indirect}) can be skipped whenever the all-variable adjacency matrix constructed from matrix $G^{\alpha\beta}$ (step $i)$ of the algorithm) does not display any link between variables in $\mathcal{A}$ and variables in $\mathcal{B}$, which is a necessary condition to have a direct link from $\mathcal{A}$ to $\mathcal{B}$.

In the SM we validate this approach on two systems of coupled logistic maps, showing that this test allows for a consistent refinement of the community where direct links between non-consecutive communities are explicitly represented.
\vspace{0.5cm}

\emph{Appendix B: Details on test systems} --- In this section we provide details on the test systems employed in the validation tests.
To write the equations in a compact form, we use indices $\mu,\nu$ to identify different dynamical communities and  indices $\alpha,\beta$ to represent variables within the same community.
Trajectories of 10$^5$ time frames were generated for all systems.
The first $10^4$ samples of each trajectory were then discarded to eliminate equilibration artifacts.

\subsubsection{Logistic maps}
The equations of the noisy coupled logistic maps shown in Figs.~1 and 2a of the main text, structured in three communities with five variables each ($\mu,\nu\in \{0,1,2\}$; $\alpha,\beta\in\{0,1,2,3,4\}$), read:
\begin{eqnarray} 
   X_\mu^\alpha(t+1)&=X_\mu^\alpha(t) \cdot \bigg(r_\mu^\alpha-r_\mu^\alpha\cdot X_\mu^\alpha(t)
   -\sum_{\beta=0}^4 c_{\mu}^{\beta\alpha} X_\mu^\beta(t) \nonumber \\
   &-\sum_{\nu=0}^2 d_{\nu\mu}^{\alpha} X_\nu^\alpha(t) + \sigma \mathcal{R}_\mu^\alpha(t) \bigg) \text{ mod } 1\,. 
\end{eqnarray}
The terms $\sigma\mathcal{R}_\mu^\alpha(t)$, where $\sigma = 0.1$ and $\mathcal{R}_\mu^\alpha\sim \mathcal{N}(0,1)$, are independent white noises added to all variables of the system.
The coefficients $c_\mu^{\beta\alpha}$ tune the strength of the interactions within the same community, while the couplings $d_{\mu\nu}^{\alpha}$ control the interactions between different communities. 
For each $\alpha$, we chose the parameters $r^\alpha_\mu$ by sampling uniformly 10 values in $\left( 3.68,\, 4\right)$, and setting $r^\alpha_0=r^\alpha_1\neq r^\alpha_2$.
Within each community $\mu$ we considered $c_\mu^{\beta\alpha}=\delta_{\beta,\alpha-1}+0.5\,\delta_{\beta,\alpha+2}$, with conventions $x_\mu^{-1} = x_\mu^4$ and $x_\mu^{5} = x_\mu^0$, and for each variable $\alpha$ in community $\mu$ we set the interaction with variables of other communities as $d^\alpha_{\nu\mu}=0.5\left(\delta_{\nu,0}\delta_{\mu,1}+\delta_{\nu,1}\delta_{\mu,2}\right)$.
The connectivity of the system is depicted in Fig. 1 of the main text.

The complementary system displaying a direct link from the communities $\mu=0$ and $\mu=2$ (Fig.~(S5) in the SM) was constructed by considering an additional term in the inter-community interaction: $d^\alpha_{\nu\mu}=0.5\left(\delta_{\nu,0}\delta_{\mu,1}+\delta_{\nu,1}\delta_{\mu,2} + \delta_{\nu,0}\delta_{\mu,2}\right)$.

\subsubsection{Lorenz systems}
The system of 5 coupled Lorenz systems ($\mu,\nu\in\{0,1,2,3,4\}$) is described by the following Itô stochastic differential equations, reported here for a single system (or community) $\mu$:
\begin{eqnarray}\label{eq:lorenz}
    \begin{cases}
        dX^0_\mu=10\,(X^1_\mu-X^0_\mu)dt+dW^{0}_{\mu}\\
        dX^1_\mu=(X^0_\mu(28-X^2_\mu)-X^1_\mu+\\
        \phantom{dX^1_\mu=} + c \sum_{\nu=0}^4 d_{\nu\mu} (X^0_\nu)^2)dt+dW^{1}_{\mu}\\
        d{X}^2_\mu=(X^0_\mu X^1_\mu- \frac{8}{3}X^2_\mu)dt + dW^{2}_{\mu}
    \end{cases}.
\end{eqnarray}

The coupling strength was fixed to $c=0.3$ and $d_{\nu\mu}$, which defines the interaction topology among communities (Fig.~2e of the main text), was set to
\begin{equation}
    d =
    \begin{pmatrix}
    0 & 0 & 0 & 0 & 0\\
    1 & 0 & 0 & 1 & 1\\    
    1 & 0 & 0 & 0 & 0\\    
    0 & 0 & 0 & 0 & 0\\    
    0 & 0 & 0 & 0 & 0\\    
    \end{pmatrix}\,.
\end{equation}
Eqs. (\ref{eq:lorenz}) were integrated using the Euler-Maruyama algorithm for Itô equations. The trajectory was calculated with a sampling time of $\delta t = 0.003$. The noise was taken independent for each variable at each time step (autocorrelation $A(\tau)=\delta(\tau)$).

\subsubsection{Lorenz 96 systems}
The two unidirectionally coupled Lorenz 96 systems of 40 variables each ($\mu,\nu\in\{0,1\}$, $\alpha,\beta\in\{0,1,...,39\}$) are defined by the following ordinary differential equations:
\begin{eqnarray}\label{eq:l96}
    d{X}^\alpha_\mu=&&\biggl((X_\mu^{\alpha+1}-X_\mu^{\alpha-2})X_\mu^{\alpha-1}-X_\mu^\alpha+F_{\mu} + \nonumber\\
    && +  c\, \delta_{\mu,1}\, X_0^\alpha\biggr)dt + dW^\alpha_\mu,
\end{eqnarray}
where $X^{-1}_\mu=X^{39}_\mu$, $X^{40}_\mu=X^0_\mu$, $F_0=5$, $F_1=6$, and $c = 0.75$.
Eqs.~(\ref{eq:l96}) were integraterd with time step $dt = 0.03$ using uncorrelated noise and Euler-Maruyama algorithm as above.

In the Lorenz and Lorenz 96 systems, the coupling terms among communities have the same functional form employed in ref.~\cite{deltatto2024robust_inference}.

\vspace{0.5cm}

\emph{Appendix C: Validation measures}  ---
The adjusted mutual information (AMI) \cite{vinh2009information_AMI} shown in the right panels of Fig.~\ref{fig:results} is a measure of discrepancy between the dynamical communities retrieved by our algorithm, $\{\mathcal{G}_i\}$, and the ground-truth groups $\{\mathcal{G}^{gt}_i\}$.
The sets $U:=\{\mathcal{G}_i\}$ and $V:=\{\mathcal{G}^{gt}_i\}$ define two possible partitions of the $D$ dynamical variables.
We computed the AMI by using the \texttt{metrics.adjusted\_mutual\_info\_score} function in SciPy \cite{virtanen2020SciPy}, which computes it as:
\begin{equation}
    \text{AMI}(U,V) = \frac{I(U,V)- \mathbb{E}\left[I(U,V) \right]}{\left( H(U)+ H(V)\right) / 2 -\mathbb{E}\left[I(U,V) \right]}\,,
\end{equation}
where $I(U,V)$ is the mutual information between the two partitions, $H(U)$ ($H(V)$) is the Shannon entropy associated to partition $U$ (respectively $V$), and $\mathbb{E}\left[I(U,V) \right]$ is the expected mutual information between two random partitions.

To measure the agreement of the retrieved links among dynamical communities with the ground-truth connectivity, we first constructed from the final community graph an all-variable adjacency matrix 
embedding all direct and indirect links retrieved by the algorithm.
Then, we computed the accuracy of this connectivity matrix as the fraction of correctly retrieved links over the total number of links:
\begin{equation}
    \text{Accuracy} = \frac{\text{TP} + \text{TN}}{\text{P} + \text{N}}\,,
\end{equation}
where $\text{TP}$ ($\text{TN}$) is the number of true positive (negative) link detections,
and $\text{P}$ ($\text{N}$) is the total number of positive (negative) detections.

\end{document}


\preprint{APS/123-QED}

\title{Linear scaling causal discovery  from high-dimensional time series by dynamical community detection \\ (Supplemental Material)}

\author{Matteo Allione}\thanks{These two authors contributed equally to this work.}
\affiliation{Scuola Internazionale Superiore di Studi Avanzati (SISSA), Via Bonomea 265, 34136 Trieste, Italy}

\author{Vittorio Del Tatto}\thanks{These two authors contributed equally to this work.}
\affiliation{Scuola Internazionale Superiore di Studi Avanzati (SISSA), Via Bonomea 265, 34136 Trieste, Italy}

\author{Alessandro Laio}
\email{Contact author: laio@sissa.it}
\affiliation{Scuola Internazionale Superiore di Studi Avanzati (SISSA), Via Bonomea 265, 34136 Trieste, Italy}
\affiliation{International Centre for Theoretical Physics (ICTP), Strada Costiera 11, 34151 Trieste, Italy}


\maketitle
\onecolumngrid

\section{Relation with conditional independencies and constraint-based methods}
In this section we draw a connection between our framework and the language of conditional independencies employed in constraint-based methods for causal discovery, which we briefly review in the following.
We refer to refs.~\cite{runge2018causal_network_reconstruction,runge2023causal_inference} for a more comprehensive overview of this framework.

\subsection{Structural causal models and time series graphs}

As a starting point, one assumes the existence of an (unknown) ``structural causal model'' (SCM) generating the data, which consists of a set of time-discrete equations ($t\in\mathbb{Z}$) of the form
\begin{equation}\label{eq:scm}
    X^{\alpha}(t) := f^\alpha\big( pa(X^{\alpha}(t)), \eta^\alpha (t)\big)\hspace{0.5cm}(\alpha=1,...,D).
\end{equation}
In Eq.~(\ref{eq:scm}), $\{X^\alpha(t)\}$ are the observed (or endogenous) variables
, $\{f^\alpha\}$ are functions called causal mechanisms, $pa(X^\alpha(t))$ is a set containing the variables $X^\beta(t-\tau)$ ($\tau \in \mathbb{N}$) that directly cause $X^\alpha(t)$, also called parents of $X^{\alpha}(t)$,
and $\{\eta^\alpha(t)\}$ are external (or exogenous) variables, typically modeled as noise terms \cite{spirtes2001causation,Assaad}.
Here, we focus on the case in which the cause precedes its effect, although the case of contemporaneous links can also be addressed \cite{runge2020discovering_pcmciplus}.
The time series graph representing the relations in Eq.~(\ref{eq:scm}) can be constructed by drawing a link $X^\beta(t-\tau)\rightarrow X^{\alpha}(t)$ whenever $X^\beta(t-\tau)\in pa(X^\alpha(t))$.
This gives rise to an infinite directed acyclic graph (DAG).

If the noise terms are assumed to be independent of each other, one can exclude the presence of unobserved common drivers, namely external variables that are causes of two or more endogenous variables.
Given this condition, known as causal sufficiency, and the fact that the underlying graph is by construction a DAG, the joint distribution of the endogenous variables can be written in a factorized form:
\begin{equation}\label{eq:markov_factorization}
    p(\{X^\alpha(t)\}) = \prod_{\alpha, t} p\big(X^\alpha(t) \mid pa(X^\alpha(t)) \big).
\end{equation}


The causal structure of the model can be equivalently written in terms of conditional independencies, through the so-called causal Markov condition:
\begin{equation}
    X^\alpha(t)\, \independent \, X^\beta(t+\tau) \,|\, pa\left(X^\beta(t+\tau)\right) \hspace{0.5cm} (\forall\alpha,\beta=1,...,D;\,\forall t\in\mathbb{Z};\, \forall \tau >0).
\end{equation}
In words, the causal Markov condition ensures 
that, conditional on the set of all its direct causes \cite{hausman_woodward_1999}, each variable is independent of all variables which are not its effects.

In the particular case in which the SCM does not vary over time, the time series graph is an infinite repetition of the same causal patterns and we can ease the notation as:
\begin{equation}\label{eq:markov_condition}
    X^\alpha(0)\, \independent \, X^\beta(\tau) \,|\, pa\left(X^\beta(\tau)\right) \hspace{0.5cm} (\forall\alpha,\beta=1,...,D;\,\forall \tau >0).
\end{equation}
Using the graphical criterion of d-separation \cite{pearl2009causality,chicharro2014algorithms_causal_inference}, conditional independencies of the form given in Eq.~(\ref{eq:markov_condition}) can be directly inferred from the structure of the time series graph. All the methods discussed in the following paragraphs are also based on the assumption that every measurable conditional independence corresponds to d-separation among the variables in the graph.
This assumption is known as faithfulness \cite{spirtes2001causation,runge2018causal_network_reconstruction}.
The Markov condition and the faithfulness assumption ensure that the structure of Eqs.~(\ref{eq:scm}) can be fully characterized in terms of conditional independencies \cite{geiger1990identifying,verma1990causal}.
Such independencies are commonly inferred from data using conditional mutual information or partial correlation.

Under the assumptions specified above, causal discovery methods that aim to infer the specific lags of the interactions for each pair of nodes $X^\alpha(t)$ and $X^\beta(t+\tau)$ will be here referred to as ``lag-specific''.
In contrast, we call ``lag-unspecific'' those methods that only aim to understand if a link between $X^\alpha(t)$ and $X^\beta(t+\tau)$ is present for any $\tau$.

\subsection{Lag-specific methods - Full conditioning}

We now discuss two strategies that can be adopted to find the parents of each node in the causal graph.

The first strategy consists in performing independence tests among a variable $X^\beta(\tau)$ and a variable $X^\alpha(0)$, while conditioning on the full history of the trajectory up to time $\tau$, denoted as $\boldsymbol{X}(\tau^{-}):= (\boldsymbol{X}(\tau-1),\boldsymbol{X}(\tau-2),...)$, excluding $X^\alpha(0)$. It can be easily shown that independence is found only if $X^\alpha(0)$ is not among the parents of $X^\beta(\tau)$.
This strategy, called ``Full Conditional Independence'' (FullCI) by Runge \emph{et al.} \cite{runge2018causal_network_reconstruction,runge2019detecting_quantifying}, corresponds to case (a) in Table~\ref{tab:comparison} and Fig.~\ref{fig:comparison}.
In practical implementations, the past of $X^\beta(\tau)$ is included up to a maximum time lag $\tau_{\text{max}}$, which is supposed to be larger than the maximum lag among all direct links \cite{runge2023causal_inference}.
This method is algorithmically efficient, as it requires a single conditional independence test between each pair of variables, but is affected by a high rate of false negative detections \cite{runge2019detecting_quantifying}.
The reason for this drawback is that, when $X^\alpha(0)\in pa\left(X^\beta(\tau)\right)$, $\boldsymbol{X}(\tau^{-})$ is likely to contain many variables that, although not necessary to assess whether $X^\alpha(0)\rightarrow X^\beta(\tau)$, explain part of the dependence between $X^\alpha(0)$ and $X^\beta(\tau)$.

\subsection{Lag-specific methods - Optimal conditioning}\label{sec:lagspecific_optimalcond}

Another possible strategy 
is to search, among all subsets of $\boldsymbol{X}(\tau^{-})$, for a subset $S$ of increasing dimension for which $X^\alpha(0)\, \independent \, X^\beta(\tau) \,|\, S \setminus X^\alpha(0)$. 
If no such subset exists, then $X^\alpha(0)$ must be a parent of $X^\beta(\tau)$.
When this occurs, the search for $S$ does not stop until all conditioning sets have been tested.
If all possible sets are considered, this results in a combinatorial explosion in the number of tests to be performed.
In practical applications, $\boldsymbol{X}(\tau^{-})$ is constructed by including all variables up to a maximum time lag in the past.
This second class of approaches includes standard algorithms such as PC \cite{spirtes1991analgorithm_PC,spirtes2001causation} or IC \cite{verma1990equivalence_IC,pearl2009causality}, designed for time-independent causal graphs, and modern approaches developed for time series data, such as PCMCI \cite{runge2019detecting_quantifying} and PCMCI+ \cite{runge2020discovering_pcmciplus}, which employ strategies to reduce the dimension of the search space.
We consider PCMCI as representative of this class of methods (see case (b) in Table~\ref{tab:comparison} and Fig.~\ref{fig:comparison}).
PCMCI can detect the presence of conditional dependencies building at each step conditioning sets of minimal size, without introducing variables that may increase the false negative rate of conditional dependence measures.

\subsection{Lag-unspecific methods - Multivariate Granger Causality / Transfer Entropy}

\begin{figure*}[ht!]
    \centering
    \includegraphics[width=17cm]{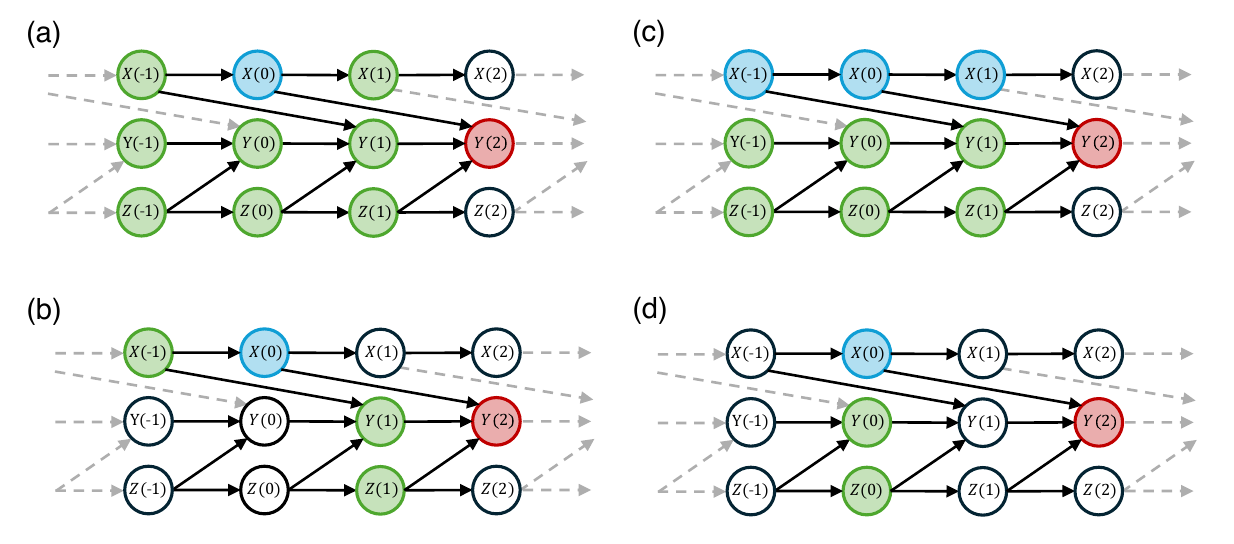}
    \caption{Visualization of different conditioning approaches on a time series graph: (a) FullCI \cite{runge2018causal_network_reconstruction}, (b) PCMCI \cite{runge2019detecting_quantifying}, (c) multivariate Granger Causality \cite{granger1969investigating,barrett2010multivariate_GC,blinowska2004granger} / Transfer Entropy \cite{schreiber2000measuring}, and (d) our approach.
    (a) and (b) are lag-specific strategies, while (c) and (d) are lag-unspecific.
    }
    \label{fig:comparison}
\end{figure*}

Multivariate Granger Causality (GC) \cite{granger1969investigating,barrett2010multivariate_GC,blinowska2004granger} and Transfer Entropy (TE) \cite{schreiber2000measuring} can be seen as lag-unspecific versions of the FullCI approach, where $X^\alpha(0)$ is replaced by the full history of $X^\alpha$ up to $t=\tau-1$, denoted by $\boldsymbol{X}^\alpha(\tau^{-})$ (see case (c) in Table~\ref{tab:comparison} and Fig.~\ref{fig:comparison}).
Also in this case, $\boldsymbol{X}^\alpha(\tau^{-})$ is constructed with frames up to a maximum time lag $\tau_{\text{max}}$ in the past.
In the case of multivariate GC, the conditional independence test is practically implemented by fitting separately two vector autoregressive models for $X^\beta(\tau)$, one including the whole past $\boldsymbol{X}(\tau^{-})$, and one including only $\boldsymbol{X}(\tau^{-}) \setminus \boldsymbol{X}^\alpha(\tau^{-})$.
Here, lag-unspecific means that condition $\boldsymbol{X}^\alpha(\tau^{-})\, \independent\, X^\beta(\tau) \,|\, \boldsymbol{X}(\tau^{-}) \setminus \boldsymbol{X}^\alpha(\tau^{-})$ allows inferring the existence of at least a direct link $X^\alpha(0)\rightarrow X^\beta(\tau)$, but not the specific value of $\tau$.
The advantages and drawback of this framework are similar to those of FullCI.
\begin{table*}[hb!]
\tabcolsep=0pt
\begin{tabular*}{14cm}{@{\extracolsep{\fill}}lcc@{\extracolsep{\fill}}}
\toprule%
& Conditioning test & Type  \\ \hline
 (a) FullCI \cite{runge2018causal_network_reconstruction}  & $X^\alpha(0)\, \independent\, X^\beta(\tau) \,|\, \boldsymbol{X}(\tau^{-}) \setminus X^\alpha(0)$ & lag-specific  \\ 
 (b) PCMCI \cite{runge2019detecting_quantifying} & $X^\alpha(0)\, \independent\, X^\beta(\tau) \,|\, \hat{pa}\left(X^\alpha(0) \right) \cup \hat{pa}\left( X^\beta(\tau)\right)$ & lag-specific  \\
 (c) Multivariate GC \cite{granger1969investigating,barrett2010multivariate_GC,blinowska2004granger} / TE \cite{schreiber2000measuring} & $\boldsymbol{X}^\alpha(\tau^{-})\, \independent\, X^\beta(\tau) \,|\, \boldsymbol{X}(\tau^{-}) \setminus \boldsymbol{X}^\alpha(\tau^{-})$ & lag-unspecific  \\ 
 (d) Our approach & $X^\alpha(0)\, \independent\, X^\beta(\tau) \,|\, \boldsymbol{X}(0) \setminus X^\alpha(0)$ & lag-unspecific  \\
 \hline
\end{tabular*}
\caption{\label{tab:comparison}Comparison of different conditioning strategies for time series causal discovery. 
The independence relationship shown in ``conditioning test'' corresponds to the null hypothesis of $X^\alpha$ being non-causal to $X^\beta$, either in a (a)-(b) lag-specific or (c)-(d) lag-unspecific fashion.
In case (b), $\hat{pa}\left(X^\alpha(0) \right)$ and $\hat{pa}\left( X^\beta(\tau)\right)$ denote the inferred set of parents of $X^\alpha(0)$ and $X^\beta(\tau)$, as the ground-truth sets are unknown.}
\vspace{1.2cm}
\end{table*}

\subsection{Connection between DII approach and conditional independencies}
By employing the DII as a metric for conditional independence, we can convert causal discovery into an optimization problem. 

In ref.~\cite{deltatto2024robust_inference} we employed Eq.~(2) of the main text,
\begin{equation}
    \hat{w} = \argmin_w \Delta\big(d^{w X(0),Y(0)}\rightarrow d^{Y\left(\tau\right)} \big) \neq 0\,
\end{equation}
as a condition to assess whether $X(0)\, \independent\, Y(\tau) \,|\, Y(0)$.
Similarly, in the generalization considered in this work,
\begin{equation}
    \hat{\boldsymbol{w}}_{\beta}
    =\argmin_{\boldsymbol{w}} \ \Delta\left(  d^{\boldsymbol{w}\odot \boldsymbol{X}(0)} \rightarrow d^{X^\beta(\tau)} \right)\,, 
\end{equation}
we considered
$\hat{{w}}_{\beta}^\alpha \neq 0$ - where $\hat{{w}}_{\beta}^\alpha$ denotes the $\alpha$-component of $\hat{\boldsymbol{w}}_{\beta}$ -  to assess
the conditional independence relationship $X^\alpha(0)\, \independent\, X^\beta(\tau) \,|\, \boldsymbol{X}(0) \setminus X^\alpha(0) $.
As in GC-inspired approaches, this method enables the identification of multi-body interactions, characterized by multiple components of \(\boldsymbol{w}\) being nonzero simultaneously.
In supplementary section~\ref{sec:windows} we show how our approach can be extended to include consecutive time frames in the optimized distance space, resulting in a practical implementation of the conditioning test $\boldsymbol{X}^\alpha(0^{-})\, \independent\, X^\beta(\tau) \,|\, \boldsymbol{X}(0^{-}) \setminus \boldsymbol{X}^\alpha(0^{-}) $.

\begin{figure*}[ht!]
    \centering
    \includegraphics[width=16cm]{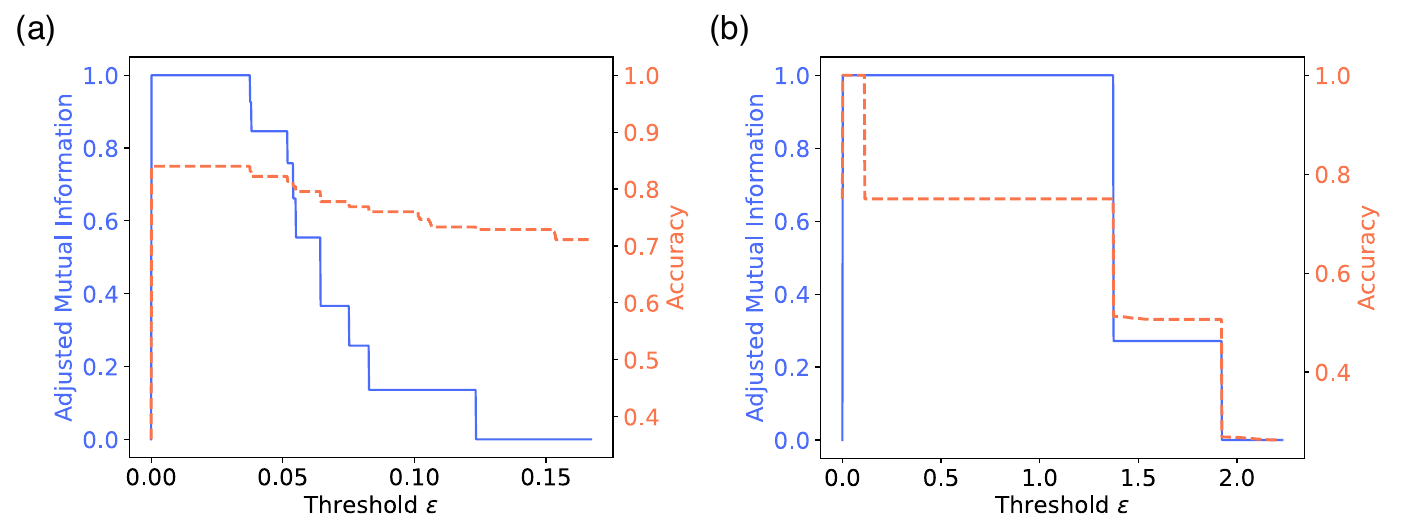}
    \caption{Validation measures referred to (a) the five coupled Lorenz systems and (b) the two coupled Lorenz 96 systems, computing the connectivity matrix $G$ with $\tau_{\text{max}} = 1$.
    In case (a), no choice of the threshold parameter $\varepsilon$ allows for a correct reconstruction of the community graph.}
    \label{fig:tau1}
\end{figure*}

Even with this extension, a key distinction remains between our conditioning strategy (case (d) in Table~\ref{tab:comparison} and Fig.~\ref{fig:comparison}) and that employed in multivariate GC / TE. Specifically, we allow for time lags $\tau\neq 1$ between the conditioning set in the past and the target variable in the future, without including the intermediate frames in the conditioning.
We observed that this is particularly relevant when analyzing time series generated by time-continuous processes. Indeed, in these cases the conditional dependence between $X^\alpha(0)$ and $X^\beta(\tau)$ can be more easily detected by our approach for $\tau > 1$, even though the ground-truth interaction occurs at shorter time scales.
As shown in Fig.~\ref{fig:tau1}, using $\tau = 1$, which is equivalent to condition up to the immediate past of the target variable, can significantly degrade the reconstruction power of our algorithm.
The parameter $\tau$ has no direct counterpart in standard approaches for causal graph reconstruction, aside from the conditional mutual information introduced by \citet{palus2001synchronization, palus2007directionality_SI}. 

We highlight that the use of $\tau\neq 1$ is only possible in a lag-unspecific framework as that considered here, where no distinction between direct and indirect causes is made.



\section{PCMCI results for the Lorenz 96 system}
As discussed in the main text, we applied the PCMCI algorithm \cite{runge2019detecting_quantifying}, as implemented in the Python library Tigramite \cite{tigramite}, to reconstruct the group structure of the Lorenz 96 systems.
We recall that PCMCI is a ``lag-specific'' method designed to reconstruct the all-variable time series graph (see supplementary section~\ref{sec:lagspecific_optimalcond}).
To enable a meaningful comparison with our approach, we employed PCMCI in a ``lag-unspecific'' fashion, ignoring unoriented links and the precise lag associated with the identified connections, as given by the \texttt{graph} output of the algorithm.
In practice, we constructed a connectivity matrix analogous to our matrix $G$ (Eq.~(4) in the main text). We considered two variables to be linked only if a direct link, denoted by entry \texttt{'-->'} in the retrieved graph matrix, was detected for any time-lag. 
Then, as in our approach, we reconstructed the dynamical communities and their links.

A trajectory of 2000 time steps was used, and conditional independence tests were conducted using the CMIknn estimator \cite{pmlr-v84-runge18a} with \texttt{model\_selection\_folds}$=3$.
First, we tested the algorithm with several combinations of the following hyper parameters: \texttt{tau\_max}, which defines the maximum time lag of direct links, \texttt{pc\_alpha}, namely the the significance threshold in the algorithm (which is automatically optimized using model selection criteria), \texttt{max\_conds\_dim}, which is the maximum number of conditions to test, and \texttt{alpha\_level}, the significance level used to construct the output graph matrix.
As an additional analysis, we applied the algorithm to a subsampled version of the trajectory with different strides.
The rationale behind this approach was that weaker dependencies might become more apparent over larger time distances and could potentially be easier to detect in this setup, as was the case with our algorithm. 

We started by carrying out a grid search for \texttt{pc\_alpha}$\,\in\{0.4, 02, 0.1,0.05\}$ and \texttt{max\_conds\_dim}$\,\in\{\texttt{None}, 5, 20\}$ (if \texttt{None}, the search is unrestricted), \texttt{alpha\_level}$\,\in\{0.01,0.04,0.05,0.07,0.1\}$ and stride$\,\in\{1,10,20,30,40\}$, obtaining in the best case 
a perfect reconstruction of only one of the 2 groups. 
In this first phase \texttt{tau\_max}$=20$ was chosen; the remaining options of the algorithm, including \texttt{tau\_min}, \texttt{max\_combinations}, \texttt{max\_conds\_px}, \texttt{max\_conds\_py} and \texttt{fdr\_method}, were set to their default values \cite{tigramite}. 
Finally, we focused on stride$\,\in\{1,5\}$ and manually fine-tuned the different parameters of the algorithm, identifying in both cases an optimal combination that resulted in the correct community causal graph.



In Fig.~\ref{fig:PCMCI_fig}a, we show the adjusted mutual information and accuracy measures (see \emph{Appendix C}) for all tested combinations of hyper parameters which gave results within 12hrs of computations.
In panels b anc c we report the connectivity matrices $G$, obtained as described above, for the two optimal hyper parameter combinations (``1'' entries are shown in yellow).
As observed from these matrices, only few dependencies among the two groups are detected.
Panels d and e display the corresponding all-variables graphs.



\begin{figure*}[ht!]
    \centering
    \includegraphics[width=\textwidth]{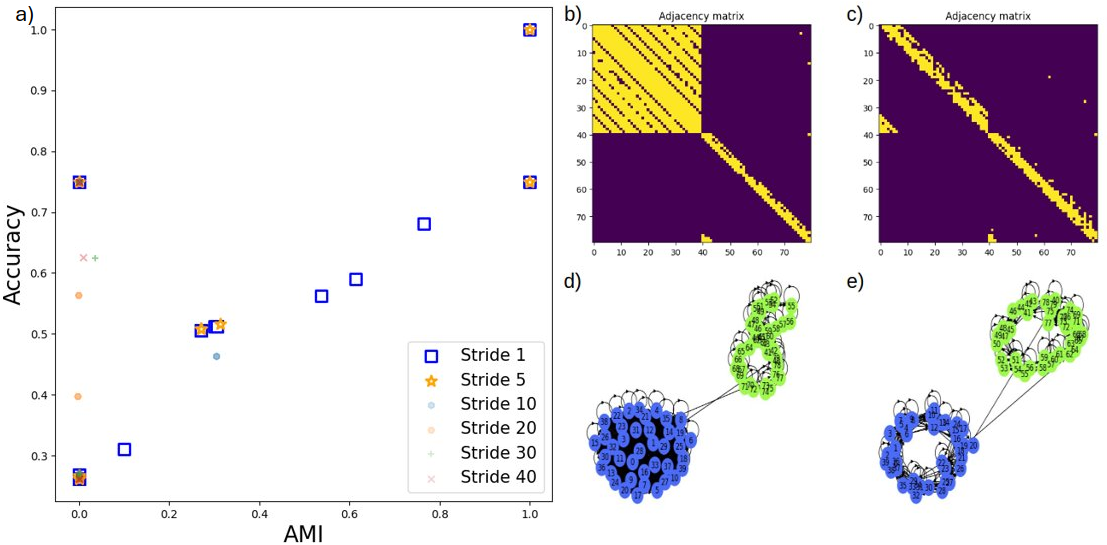}
    \caption{(a) Scatter plot showing Adjusted Mutual Information (AMI) and accuracy values for different hyperparameter settings. 
    Among the tested combinations, only a few achieved perfect community detection (AMI = 1). The link between the communities was challenging to identify. For stride$=1$, we found that \texttt{tau\_max}$=20$, \texttt{alpha\_level}$=0.00625$, \texttt{max\_conds\_dim}$=20$ allows for a perfect reconstruction (top right point); for stride$=5$, the same was obtained for \texttt{tau\_max}$=6$, \texttt{alpha\_level}$=0.015$, \texttt{max\_conds\_dim}$=4$. 
    For the best-performing configuration we display on the right the retrieved adjacency matrix and reconstructed graph. Panels b) and d) refers to  stride$=1$, c) and e) to  stride$=5$.
    }
    \label{fig:PCMCI_fig}
\end{figure*}

\section{Generalization to time windows}\label{sec:windows}
In the main text, we presented our approach constructing distance $d^A$ with all variables at a single time frame ($t=0$), assuming that the maximum lag of direct links is not larger than 1 in the underlying time series graph.
Here we show how the conditioning should be extended to multiple frames when this assumption does not hold, and how this can be simply achieved by constructing $d^A$ on a time window rather than on a single time frame.

We use as illustrative example the following three coupled logistic maps:
\begin{eqnarray}\label{eq:logistic_confounder}
    \begin{cases}
        X(t+1)=X(t) \cdot \left(r_X-r_X\cdot X(t) + \sigma \mathcal{R}_X(t) \right) \text{ mod } 1\\
        Y(t+1)=Y(t) \cdot \left(r_Y-r_Y\cdot Y(t) - c \cdot X(t-1) + \sigma \mathcal{R}_Y(t) \right) \text{ mod } 1\\
        Z(t+1)=Z(t) \cdot \left(r_Z-r_Z\cdot Z(t) - c \cdot X(t)  + \sigma \mathcal{R}_Z(t) \right) \text{ mod } 1
    \end{cases}.
\end{eqnarray}
The terms $\sigma\mathcal{R}_\alpha(t)$ ($\alpha=X,Y,Z$), where $\sigma = 0.1$ and $\mathcal{R}_\mu^\alpha\sim \mathcal{N}(0,1)$ are three independent white noises.
The parameters $r_X$, $r_Y$, and $r_Z$ were randomly sampled in the interval $\left(3.68,4\right)$, resulting in $r_X \simeq 3.863$, $r_Y \simeq 3.861$ and $r_Z \simeq 3.836$.
As for the other systems, a time series of 10$^5$ samples was generated, discarding the first 10$^4$ initial points and then sampling $N=2000$ independent initial conditions for the DII optimization.

\begin{figure*}[ht!]
    \centering
    \includegraphics[width=\textwidth]{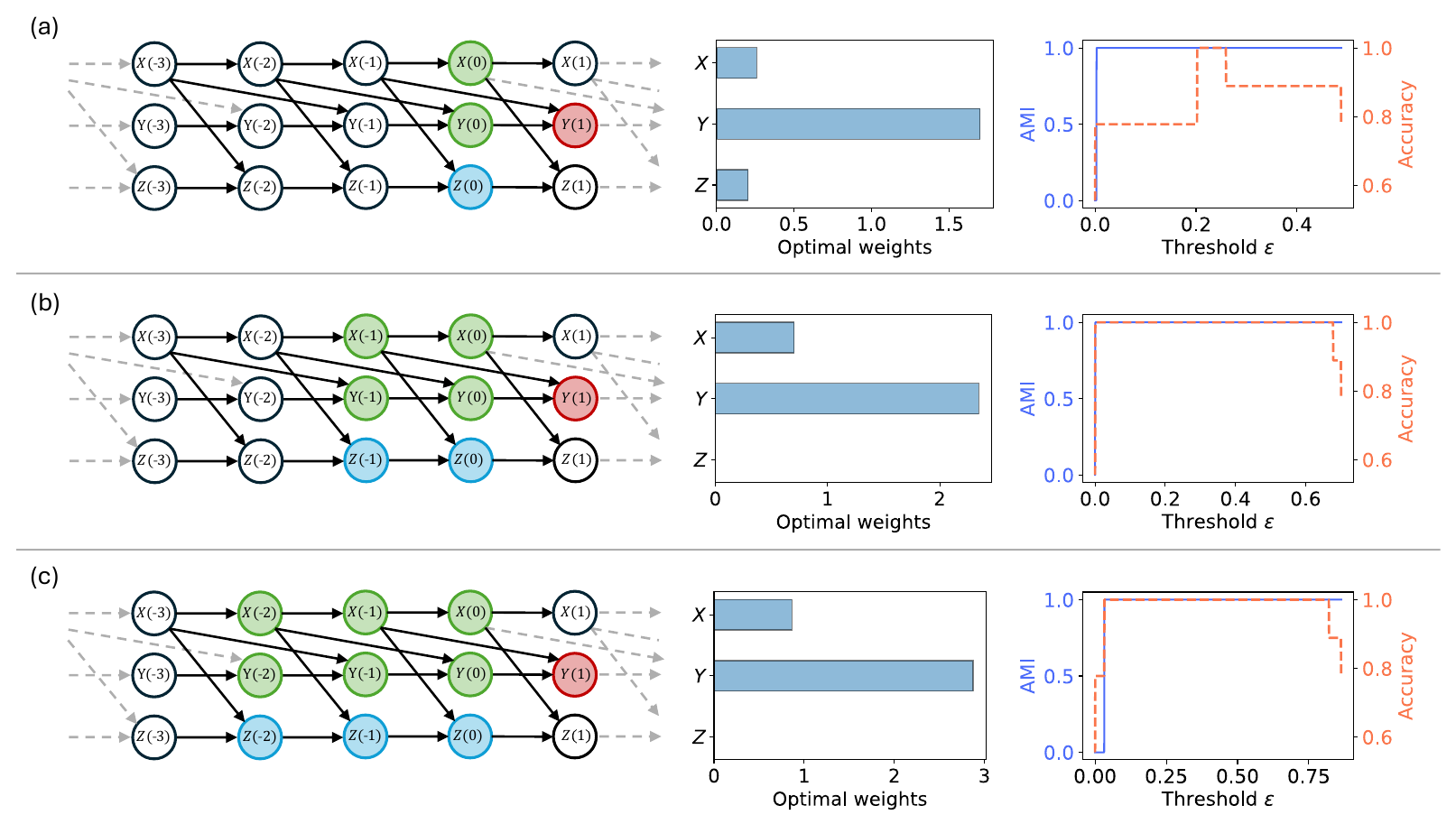}
    \caption{Extension of the method to time windows, using three noisy coupled logistic maps.
    The time series graphs in (a), (b) and (c) show the construction of the distance $d^A$ over different number of frames ($E=1,2$ and 3, respectively), including all variables represented by green and blue nodes.
    The central bar plots show the optimal DII weights using as target variable $Y(1)$ (red node in the time series graphs).
    When $E > 1$, only the largest weight across different frames is shown for each variable.
    On the right, the adjusted mutual information (AMI) and the accuracy of the retrieved connectivity matrix are plotted as a function of the threshold parameter $\varepsilon$, in the range $\left[0, \max_{\alpha\beta\,(\alpha\neq\beta)}G^{\alpha\beta} \right]$.
    }
    \label{fig:windows}
\end{figure*}

According to Eqs.~(\ref{eq:logistic_confounder}), $X$ causes $Y$ with lag 2 and $X$ causes $Z$ with lag 1.
The time series graph of the process is shown in Fig.~\ref{fig:windows}.
In this example, the dynamical communities are simply $\{X \}$, $\{Y \}$ and $\{Z \}$, and the community graph is $\{Z \}\leftarrow \{X\}\rightarrow \{Y \}$.
Hereafter we will focus on $Y(\tau=1)$ as target variable of the DII optimization (red node in the time series graphs of Fig.~\ref{fig:windows}).

Fig.~\ref{fig:windows}a shows the application of our approach according to Eq.~(5) in the main text, namely minimizing $\text{DII}\left(  d^{\boldsymbol{w}\odot \boldsymbol{X}(0)} \rightarrow d^{Y(\tau=1)} \right)$ with $\boldsymbol{X}(0) = (X(0),Y(0),Z(0))$.
In this case, a spurious weight associated to $Z(0)$ appears due to the presence of an open path $Z(0)\leftarrow X(-1) \rightarrow Y(1)$.
The term ``open'' and ``closed'' (or ``blocked'') are used here in the sense of d-separation \cite{chicharro2014algorithms_causal_inference, runge2018causal_network_reconstruction}: two variables in the graph are conditionally dependent when they are connected by at least an open path.
The non-zero weight associated to $Z(0)$ affects the reconstruction quality of the algorithm, resulting in a limited range of the threshold $\varepsilon$ providing the correct community graph.

To overcome this problem, we can construct the first distance space as $d^{\boldsymbol{w}\odot\boldsymbol{X}_E(0^{-})}$, where $\boldsymbol{X}_E(0^{-})$ is the time window of $E$ frames including all variables from $t=-E+1$ to $t=0$, namely
\begin{equation}
    \boldsymbol{X}_E(0^{-}) = \bigg( \underbrace{X(0), Y(0), Z(0)}_{\boldsymbol{X}(0)}, \underbrace{X(-1), Y(-1), Z(-1)}_{\boldsymbol{X}(-1)}, ..., \underbrace{X(-E+1)), Y(-E+1), Z(-E+1)}_{\boldsymbol{X}(-E+1)}\bigg)\,.
\end{equation}
In turn, $\boldsymbol{w}$ denotes in this case a vector of 3$E$ components (or, more in general, $D\cdot E$ components, where $D$ is the number of variables).
Setting $E=2$ (Fig.~\ref{fig:windows}b) allows including $X(-1)$ in the conditioning set, blocking the path $Z(0)\leftarrow X(-1) \rightarrow Y(1)$.
In this case, no spurious weight associated to $Z$ appears, and the community graph reconstruction is correct for almost all values of $\varepsilon$ in the interval $\left[0, \max_{\alpha\beta\,(\alpha\neq\beta)}G^{\alpha\beta} \right]$.
The same occurs using a larger window ($E = 3$, Fig.~\ref{fig:windows}c).

In general, as can be easily demonstrated,  constructing $d^A$ on $E$ consecutive frames enables the application of our conditioning approach without errors, assuming that the maximum lag of direct links in the underlying time series graph is $E$.

\section{Direct and indirect links between dynamical communities}
In Fig.~\ref{fig:causal_graph_refinement} we validate the refinement procedure outlined in Appendix A of the main text using two complementary systems of coupled logistic maps.
Specifically, we employed the 15-dimensional system already displayed in Fig.~1 and 2a of the main text and a similar system where microscopic direct links between the community of level 0 (denoted by $\bm{X}$) and the community of level 2 (denoted by $\bm{Z}$) are added.
As a consequence, these communities are directly linked in the second system, while they are only indirectly connected in the first case.
The results obtained with our algorithm for these two systems are shown in the first and second rows of Fig.~\ref{fig:causal_graph_refinement}, respectively.
The explicit equations of both systems are reported in \emph{Appendix B}.

\begin{figure*}[ht!]
    \centering
    \includegraphics[width=\textwidth]{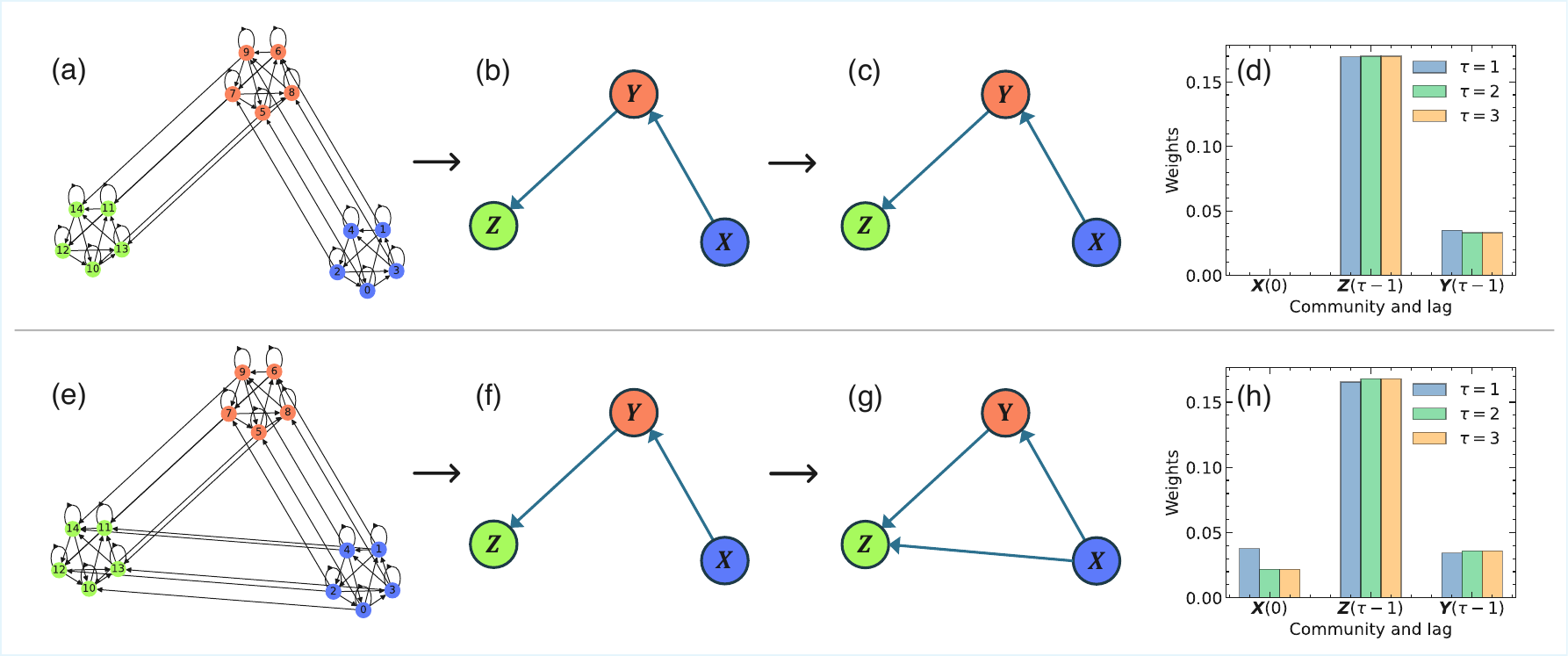}
    \caption{
    Results of the graph refinement step on two systems of coupled logistic maps.
    In the system of first row, communities $\bm{X}$ and $\bm{Z}$ are indirectly linked in the ground-truth dynamics, while they are directly linked in the system of second row.
    Panels (a) and (e): all-variable graphs obtained from condition $G_{\alpha\beta} > \varepsilon$.
    Panels (b) and (f): community graphs drawn after step \emph{iii)} of our algorithm.
    Panels (c) and (g): refined community graphs after the refinement step to identify direct links between non-consecutive communities.
    Panels (c) and (h): bar plots of the optimal weights associated to $\bm{X}(0),\, \bm{Z}(\tau-1)$ and $\bm{Y}(\tau-1)$, using $\bm{Z}(\tau)$ as target in the minimization of the DII, for different values of $\tau$.
    }
    \label{fig:causal_graph_refinement}
\end{figure*}

As panels (c) and (g) show, following the procedure in \emph{Appendix A}, a direct link between communities $\bm{X}$ and $\bm{Z}$ is correctly recovered only in the second case, for all threshold values in range $\left[0, w_{\bm{Y}(\tau -1)}
\right]$ (the upper limit of this interval represents the maximum weight associated to the community that is already known to be directly connected with $\bm{Z}$).
In such refined community graphs, direct arrows identify \emph{all} direct links between dynamical communities.
In the first system, the DII minimization used to compute the optimal weights is not strictly necessary, because the adjacency matrix from step \emph{i)} already shows no links between variables in $\bm{X}$ and variables in $\bm{Z}$ (panel (a)).
As a result, the DII optimization can be omitted in this case, improving the efficiency of the refinement procedure.

\section{Considerations on causal sufficiency \label{sec:causal_sufficiency}}
\begin{figure}
    \centering
    \includegraphics[width=0.6\linewidth]{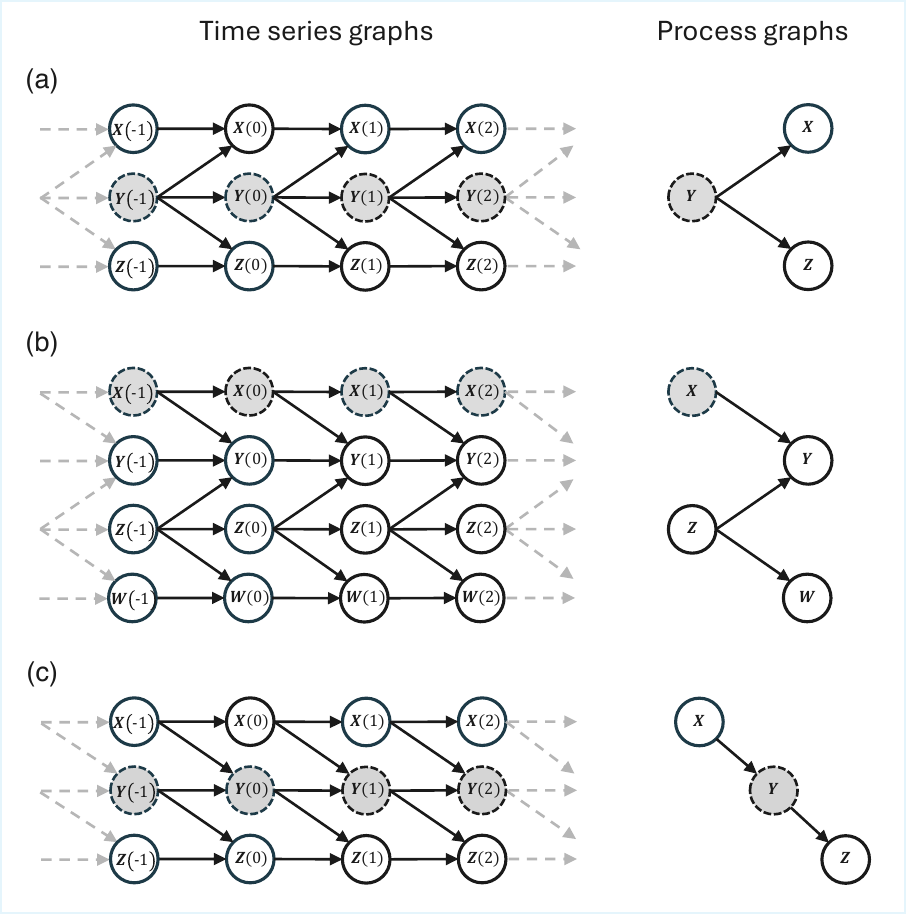}
    \caption{Examples of violations of the causal sufficiency hypothesis.
    Grey nodes with dashed edges denote unobserved variables.
    (a): $Y$ is a common driver of $X$ and $Z$.
    (b): $X$ is only causing $Y$ but has non-zero autocorrelation.
    (c): $Y$ mediates the interaction from $X$ to $Y$.}
    \label{fig:causal_suff}
\end{figure}

The method described in the main text relies on the assumption of causal sufficiency, which excludes the existence of latent (unobserved) variables causing two or more endogenous (observed) variables.
In a microscopic time series graph, each dynamic variable at a given time can be seen as a distinct endogenous variable, which is therefore represented by a single node.
In this section we will call ``process'' the collection of all nodes describing the time evolution of a given dynamic variable.
A dynamic process is represented as a single node in a process (or summary) graph~\cite{reiter2023CausalIO}.
In Fig.~(\ref{fig:causal_suff}), we support our discussion using different examples in terms of time series and process graphs.
In such examples, all dynamical communities are composed by single variables; therefore, the ground-truth community graphs are trivially equivalent to the process graphs.

In terms of processes, the causal sufficiency assumption is violated when a process acting as a common driver of two or more observed processes is excluded from the analysis (panel a), or when an unobserved and autocorrelated process is the cause of a single endogenous process.
The latter scenario may occur when the unobserved process is exogenous, namely not caused by any observed process (panel b), or when it is a mediator of two observed processes (panel c).
For simplicity, we exclude from our discussion the case of an unobserved mediator process being directly caused by two or more observed processes and the presence of multiple unobserved processes.

In the case depicted in panel a, the presence of an unobserved common driver brings to the detection of a spurious bidirectional link between $X$ and $Z$.
In fact, one can observe by applying the rules of $d$-separation~\cite{pearl2009causality,chicharro2014algorithms_causal_inference} that $X(1)$ is not $d$-separated from $Z(0)$ given the conditioning set $\{ X(0)\}$, due to the presence of the open path $Z(0)\leftarrow Y(-1) \rightarrow Y(0) \rightarrow X(1)$. 
Similarly, $Z(1)$ is not $d$-separated from $X(0)$ given the conditioning set $\{ Z(0)\}$.
As a consequence, our algorithm would find the autonomous sets $S^X = S^Z = \{X,Z \}$ if $Y$ was unobserved, resulting in a community graph with a single dynamical community. 
Therefore, because of the presence of the latent variable $Y$, $X$ and $Z$ gets aggregated into a single node instead of being identified as separated communities.
Importantly, this spurious detection cannot be avoided by including previous time frames in the conditioning set.

In the system of Fig.~\ref{fig:causal_suff}b, 
the presence of an autocorrelated process $X$ which is coupled to $Y$ results in a dynamics that is not anymore Markovian in the subsystem of processes $Y$, $Z$ and $W$, when $X$ is unobserved.
In practice, this is undistinguishable from a dynamics in the variables $Y$, $Z$ and $W$ which displays direct ``self-links'' $Y(0)\rightarrow Y(\tau)$ extending beyond $\tau = 1$.
Therefore, assuming that the maximum lag of the direct links is 1 is not appropriate in the subsystem of the observed variables, and may result in the detection of spurious connections.
For example, one may observe that $Y(1)$ is not d-separated from $W(0)$ given the conditioning set $\{X(0),\,Y(0),\,Z(0) \}$, because the path $W(0)\leftarrow Z(-1) \rightarrow Y(0) \leftarrow X(-1) \rightarrow X(0) \rightarrow Y(1)$ is open.
In this example, using a single time frame at $t=0$ would bring to the autonomous sets $S^Y =\{Y,Z,W\}$ (rather than $S^Y =\{Y,Z\}$), $S^Z = \{Z\}$ and $S^W = \{Z,W\}$, resulting in a community graph with a spurious link between communities $\{W\}$ and $\{Y\}$.
However, our method can still provide the correct community graph by replacing the single frame at time $t=0$ with the entire time series history, or by using time windows that are sufficiently long for self-dependencies to effectively decay (see SM section~\ref{sec:windows}).
Notice that, for example, the path above is blocked when the conditioning set is extended to $\{X(0),\,Y(0),\,Z(0),\,X(-1),\,Y(-1),\,Z(-1)  \}$.

Finally, in the example of Fig.~\ref{fig:causal_suff}c the outcome of our algorithm is unaffected, as the presence of unobserved mediators between two observed processes 
cannot invert the directionality of the information flow between the variables.
On the contrary, this case is problematic for methods targeting the microscopic time series graph, as direct links may be identified in place of indirect connections, and wrong time lags may be inferred \cite{runge2018causal_network_reconstruction}.


In conclusion, when the assumption of causal sufficiency is violated, we can expect two main types of artifacts in the inferred structure: some detected communities may in fact represent the union of multiple true communities, merged due to the influence of unobserved common drivers, and direct links between communities might be spurious, if the method is applied using time windows of insufficient length.

\section{Observational noise}
\begin{figure}[h]
    \centering
    \includegraphics[width=0.8\linewidth]{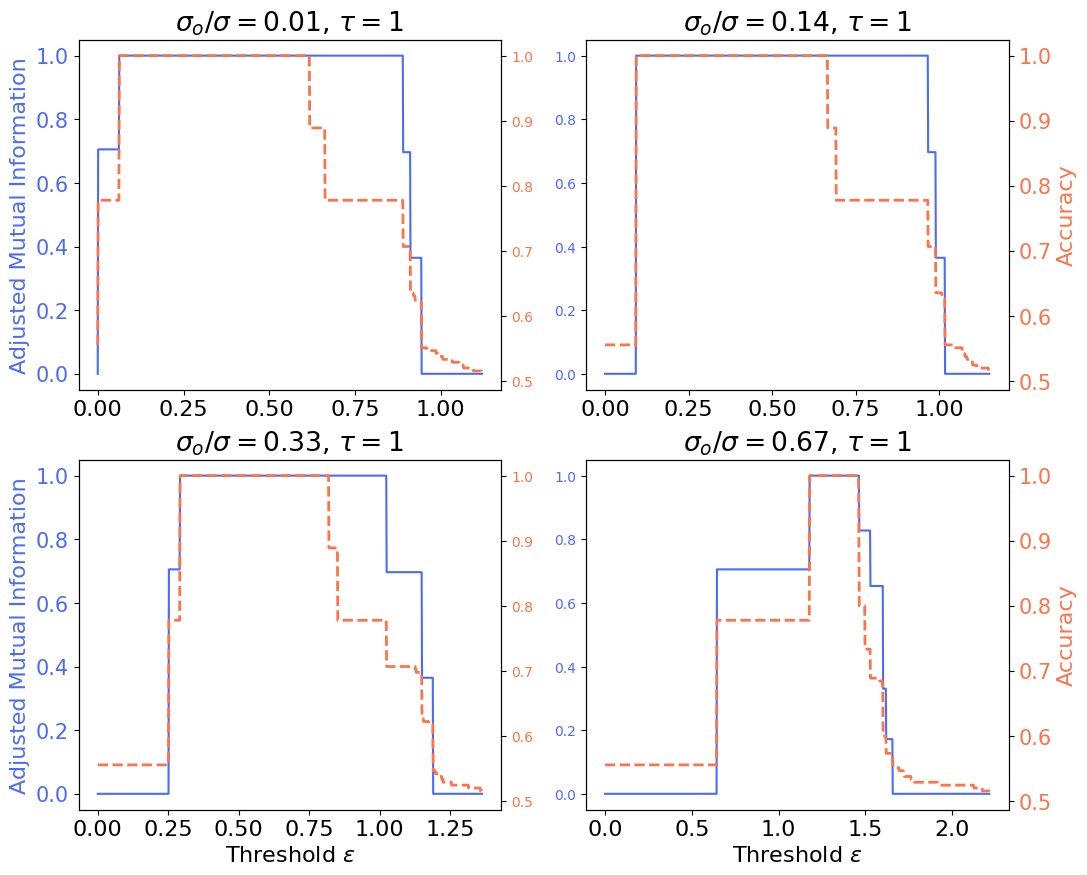}
    \caption{Values of the Adjusted Mutual Information (blue) and Accuracy (orange) for different levels of observational noise, expressed in terms of the ratio between the standard deviations of the observed noise $\sigma_o$ and the standard deviation of the variables of the system $\sigma$.
    Results are shown by varying the threshold $\varepsilon$ in the range $\left[0, \max_{\alpha \beta\, (\alpha\neq \beta)} G^{\alpha\beta}\right]$.}
    \label{fig:obs_noise}
\end{figure}

We analysed the effect of observational noise on the reconstruction of the community graph associated to the logistic maps of Eq.~(9) in \emph{Appendix B}. 
Given the noiseless trajectory $X^\alpha_\mu(t)$, we injected additive noise in the form
\begin{equation}\tilde{X}^\alpha_\mu(t)=X^\alpha_\mu(t)+\tilde{\mathcal{R}}_\mu^\alpha(t),   
\end{equation}
where 
$\tilde{\mathcal{R}}_\mu^\alpha(t)$ are independent Gaussian white noises, whose standard deviation was set to a fraction $f$ of the empirical standard deviation of $X^\alpha_\mu$.
In Fig.~\ref{fig:obs_noise} we show the AMI and the accuracy of the retrieved community graph, as a function of the threshold $\epsilon$, for $f = 0.01,\,0.14,\,0.33$ and 0.67.
The performance of the algorithm is almost unaffected up to $f = 0.14$, although an exact reconstruction is still achievable for significantly larger noise magnitudes.
As a main effect of observational noise, we observe an increase of the critical lower bound threshold that achieves zero false positives (shown by the first value at which AMI and accuracy become 1 in the panels of Fig.~\ref{fig:obs_noise}).


\section{Details on the DII optimization}
In this section, we provide details on the DII optimization process.
The code, available in the Python package DADApy \cite{dadapy2022}, was implemented in JAX \cite{jax2018github}.

\subsection{Neighborhood size parameter $\lambda$}

An important parameter appearing in Eq.~(4) of the main text is $\lambda$, which defines the size of the neighbourhoods in the first distance space.
The classical Information Imbalance is recovered in the limit $\lambda \to 0$; however, excessively small $\lambda$ values can hinder the optimization process, as the DII derivatives approach zero in this regime \cite{wild2025automatic}.
To address this issue, we employed a point-adaptive scheme for computing $\lambda$, assigning a distinct value $\lambda_i$ to each point $i$ in the sum of Eq.~(4) in the main text.
In this formulation, the standard Information Imbalance is recovered in the limit $\lambda_i\rightarrow 0$, $\forall i =1, ..., N$.
The use of a point-specific $\lambda_i$ is particularly beneficial when the data set features a non-homogeneous point distribution.
In this case, relying on a single distance scale could result in inconsistent neighborhood sizes across different regions of the data manifold.
Each $\lambda_i$ was computed as
\begin{equation}\label{eq:lambda_eq}
    \lambda_i = 0.1\, d^{A}_{ij(k)}(\boldsymbol{w})\,,
\end{equation}
where 0.1 is an empirical prefactor and $j(k)$ is the $k$-th nearest neighbour of $i$ according to the (squared) distance $d^{A}(\boldsymbol{w})$.
In this study, $k$ was fixed to 5\% of the total number of points used in the DII calculation (namely, $k/N = 0.05$). 
To account for changes in $d^{A}(\boldsymbol{w})$ during the optimization, the parameters ${\lambda_i}$ were recomputed after each weight update.

\subsection{Optimization strategy}
The convergence of the DII optimization to its global minimum depends on several factors, including the choice of the optimizer and the use of mini-batches. 
Mini-batching, which involves computing gradients on random subsets of points during each gradient descent update, is a common strategy to improve both convergence speed and stability, particularly when the loss function contains multiple local minima.

In this work, we sampled $N = 2000$ evenly spaced frames from each time series.
In the DII optimization, for each training epoch, we randomly partitioned the resulting data set into 20 mini-batches, each containing $N' = 100$ points.
Within each mini-batch, the parameter $k$ for determining ${\lambda_i}$ was set to 5.
The DII optimization was carried out using the Adam optimizer \cite{kingma2017adam}, which is well known for its robust convergence properties.

\subsection{Training Schedule}
In this work, all DII optimization were carried with 500 training epochs.
In each optimization, the learning rate used by the Adam optimizer was set to the initial value of $5\times 10^{-3}$, and gradually decreased to zero according to a cosine decay schedule.

\bibliography{suppl_mat}